\documentclass[manuscript]{aastex}

\begin{document}
\pagenumbering{arabic}
\title{HOW DIFFERENT ARE NORMAL AND BARRED SPIRALS?}
\author{Sidney van den Bergh}
\affil{Dominion Astrophysical Observatory, Herzberg Institute of Astrophysics, National Research Council of Canada, 5071 West Saanich Road, Victoria, BC, V9E 2E7, Canada}
\email{sidney.vandenbergh@nrc.gc.ca}

\begin{abstract}
 No significant color differences are found between normal
and barred spirals over the range of Hubble stages a - ab - b - bc.
Furthermore no significant difference is seen between the luminosity 
distributions of normal and barred galaxies over the same range of 
Hubble stages. However, SBc galaxies are found to be systematically 
fainter than Sc galaxies at 99\% confidence.  The observation that 
normal and barred spirals with Hubble stages
a - ab - b - bc
have indistinguishable intrinsic colors hints at the possibility that
the bars in such spiral galaxies might be ephemeral structures.
Finally, it is pointed out that lenticular galaxies of types S0 and
SB0 are systematically fainter than are other early-type galaxies,
suggesting that such galaxies are situated on evolutionary tracks that
differs systematically from those of galaxies that lie along the
E - Sa - Sb -Sc and E - SBa - SBb - SBc sequences.

\end{abstract}


\section{INTRODUCTION}

 The present paper asks three questions that (as far as I have been
able to determine) have not been asked before: (1) Do normal and barred 
of the same Hubble stage have different colors, (2) do normal and
barred galaxies of the same Hubble stage have different luminosities
and (3) is the frequency distribution of r (ring) and s (sprial) type substructures 
within galaxy disks different in normal and barred galaxies? It should
be emphasized that the present selection of galaxies by morphological
type is radically different from other recent investigations (e.g. 
Giordano et al. 2010, Masters et al. 2010, Nair \& Abraham 2010) that
select galaxies by color, or by luminosity and then compare the
relative frequencies of normal and barred objects.

Surprisingly there appears to be no discussion in the astronomical
literature of possible systematic differences between the colors and
luminosities of normal and barred spiral galaxies that have been 
selected on the basis of their morphologies. Since the colors of
galaxies are known to become bluer with advancing Hubble stage it is
clear that any subtle color differences between normal and barred
objects of a given Hubble stage, can only be studied effectively using
galaxy classifications that are of the very highest quality. The gold
standard of such classifications is provided by the 1246 galaxies in
{\it A Revised Shapley-Ames Catalog of Bright Galaxie}s by Sandage \& 
Tammann (1981). These classifications, by two expert morphologists,
are based exclusively on inspection of large-scale photographic plates 
that are all beautifully illustrated in {\it The Carnegie Atlas of Galaxies} 
(Sandage \& Bedke 1994). The Shapley-Ames Catalog of Bright Galaxies 
constitutes a magnitude-limited sample, which is an enormous advantage
for morphological studies. This is so because luminosity-limited samples
are biased towards luminous galaxies, which exhibit striking morphological
differences. This contrasts with the situation for volume-limited samples
which are dominated by intrinsically faint galaxies for which morphological
differences tend to be much more subtle (van den Bergh 1998, p.25). A
problem with luminosity-limited samples is, of course, that they may be
biased by luminosity selection. However, as will be seen in Section 3 of 
the present paper, there appear to be no significant luminosity differences
between normal and barred spirals over the range of Hubble stages
a - ab - b - bc. Therefore the Shapley-Ames sample is ideally suited
for a study of possible systematic differences between normal and barred
galaxies.

Edwin Hubble (1926, 1936) suggested that spiral galaxies could be arranged 
into a `tuning fork' diagram, in which the tines are represented by
normal and barred objects. Later (e. g. van den Bergh 1998) it became
clear that this dichotomy extends to the realm of the irregular galaxies
with the SMC being an example of a normal irregular an the LMC being a
barred irregular galaxy. Furthermore (see for example Sandage 1975)
lenticular galaxies were also found to exhibit either normal (S0) or
barred (SB0) morphology. Sandage (1975) and Sandage \& Tammann
(1981) assign the overwhelming majority of spiral galaxies to either the
normal or barred type, with few intermediate objects. On the other hand
de Vaucouleurs (1959) advocated a classification system in which bar
strength varied continuously from pure spirals of type SA, through
intermediate objects of type SAB, to pure bars assigned to type SB.
Galaxies that Sandage \& Tammann classified as being edge-on
were excluded from the sample because it is often difficult (or
impossible) to distinguish normal and barred spirals that are viewed
edge-on. Also excluded were those galaxies which Sandage \& Tammann
classified as being 'peculiar'. Some of such peculiar galaxies turned 
out to be unusually blue, indicating that their apparent peculiarity is 
due to (or associated with) a recent burst of star formation. The adopted 
luminosities of galaxies are the $M^{o,i}_{B_T}$ values of Sandage \& Tammann (1981). In the subsequent discussion these magnitudes will, for the sake of 
simplicity, be referred to as $M_{B}$. For the majority of the galaxies in the Shapley-Ames
catalog total (asymptotic) colors on the Johnson B-V system, that have
been corrected for Galactic and internal extinction, and for the effects of 
redshift, are available from the {\it Third Reference Catalogue of Bright Galaxies}
(=RC3) by de Vaucouleurs et al. (1991). These intrinsic colors will 
subsequently 
be designated $(B-V)_{o}$.

\section{COLORS OF NORMAL AND BARRED SPIRALS}

The following discussion is based on all those galaxies in the
Revised Shapley-Ames Catalog which are not classified as `edge-on' or as
`peculiar', and for which the RC3 catalog gives $(B-V)_o$ colors. [For both
normal spirals and for barred spirals the fraction of objects classified
as being peculiar by Sandage \& Tammann lies between 5\% and 6\%]. It
is not clear that the observed distribution of $(B-V)_{o}$ colors for any
Hubble type will be Gaussian. The following discussion is therefore
based on the median colors $(B-V)_{o}^{*}$ of different subgroups, rather
than on their mean values $<(B-V)_{o}>$. A comparison between these median
colors of normal spiral and of barred galaxies is shown in Table 1 and
is plotted in Figure 1. Also listed in this table are the $(B-V)_{o}^{*}$ colors containing 25\% and 75\% of the data points and error estimates based on the inter-quartile range divided by the square root of the number of galaxies. Inspection of the data in Table 1 shows that, within any Hubble stage, the colors of normal and barred objects are very smiliar.  This conclusion is strengthened and confirmed by Kolomogorov-Smirnov (K-S) tests which show no significant differences between the color distributions of Sa and SBa, Sab and SBab, Sb and SBb, Sbc and SBbc and Sc and SBc galaxies.  The largest 
difference in Table 1 occurs between the colors of Sc and SBc galaxies, 
with the SBc galaxies appearing, on average, slightly bluer than those of 
type Sc . The observed 
color between Sc and SBc galaxies is in the sense expected from the fact (see Section 3) that 
SBc galaxies are systematically fainter (and hence are expected to be metal poorer) 
than are objects of type Sc. It is concluded that presently available
high-quality data do not exhibit a significant difference between the
intrinsic $(B-V)_o$ colors of normal and barred galaxies over the range of
Hubble stages a - ab - b - bc.  The referee has raised the interesting question whether the presence of bars in galaxies might have affected the Hubble stage assignments of galaxies in the catalog of Sandage \& Tammann (1981) in a systematic way.  With the passing of Allan Sandage it appears unlikely that this question will ever by answered in an entirely satisfactory fashion.

\section{LUMINOSITIES OF NORMAL AND BARRED SPIRALS}

  The Revised Shapley-Ames Catalog (Sandage \& Tammann 1981) lists the
$M_{B}$ luminosities for all of the galaxies in their sample. A summary of
the data on the median luminosities of normal and barred galaxies is
given in Table 2. Also given for comparison is the median luminosity of
the elliptical galaxies listed in the Shapley-Ames catalog. These data
can be used to search for systematic luminosity differences between
normal and barred spirals. For each Hubble stage in the range a - ab -
b - bc such a comparison shows no clear-cut difference between the
luminosity distributions of normal and barred spirals. Since the
mass-to-light ratios are expected to be similar within each Hubble stage
this result suggests that normal and barred spirals of Hubble
stages a - ab - b - bc have similar masses, i.e. the difference between
barred and non-barred spirals is not a consequence of mass differences.
However, a K-S test does show (at $>99\%$ significance) that barred spiral
of type SBc are systematically fainter than normal spirals of type Sc.
For 264 Sc galaxies $M_{B}^{*}$ = -20.86, compared to  $M_{B}^{*}$ = -20.20 for 69 SBc
galaxies. Since faint galaxies of a given Hubble type are systematically
slightly bluer than luminous galaxies of the same type one would expect
the average colors of SBc galaxies to be slightly bluer than those of
type Sc. This is, in fact, the sense of the color differences listed in
Table 1. So metallicity effects may have provided a slight boost to the
apparent systematic color difference between Sc and SBc galaxies. It is
noted parenthetically that the luminosity distributions of S0 and SB0
galaxies in Sandage \& Tammann (1981) do not differ significantly. 
  The fact that the numbers of galaxies in Table 2 is greater than
that in Table 1 is due to the fact that all galaxies listed in the
Shapley-Ames catalog have been assigned luminosities, whereas B-V
colors from the RC3 are available for most, but not all, of the
individual galaxies contained in that catalog.  
  
The referee of this paper has expressed some concern about the
fact that the luminosities of field galaxies in the Shapley-Ames
catalog were determined from their radial velocities and might
therefore be affected by the random motions of galaxies. It
therefore seemed prudent to repeat the data on the median
luminosities of galaxies listed in Table 2 using only (1) galaxies 
that are members of clusters identified by Sandage \& Tammann (1981) 
and (2) field galaxies with radial velocities (corrected for
motion relative to the centroid of the Local Group) $>2000~km s^{-1}$,
for which radial velocities should provide a reasonable proxy for 
distances. The results for these galaxies for which the luminosities
are most secure are listed in Table 3. These data confirm
the previous conclusion that barred and unbarred galaxies of 
Hubble stages a - ab - b - bc  have similar luminosities, and
hence presumably comparable masses. As was the case in Table 2,
galaxies of type SBc are again (on average) found to be less luminous 
than those of type Sc.  
   
Finally a comparison between the luminosity distributions of 
various kinds of early-type galaxies is shown in Table 4 and is plotted 
in Figure 2. The most striking feature of these data is that the S0 + 
SB0 galaxies are, on average, about a magnitude fainter than are E 
and Sa + SBa galaxies. A K-S test shows that there is less than a
0.1\% chance that the S0 + SB0 sample was drawn from the same
parent population as that of the E and Sa +SBa galaxies. This result strongly 
suggests that S0 + SB0 galaxies lie on evolutionary tracks that,
on average, differ significantly from those of galaxies that fall
along the E - Sa - Sab - Sb - Sbc and E - SBa - SBab - SBb - SBbc 
sequences. The data presented above show, as has previously been
emphasized by van den Bergh (1998, p.61), that S0 galaxies are not
truly intermediate between galaxies of types E and Sa. This view
conflicts with that of Hubble (1936, pp.44-45) who introduced S0 
galaxies as a ``more or less hypothetical'' class to bridge the chasm 
between elliptical and spiral galaxies. The data listed in Table 3 
also show that the systematical luminosity difference between S0
galaxies on the one hand, and E and Sa galaxies on the other, is
also present in the sub-sample of galaxies with the best-determined
luminosities.

\section{OTHER STRUCTURE WITHIN DISKS}

   Following in the footsteps of de Vaucouleurs (1959), Sandage \&
Tammann (1981) classified the structure within some disk galaxies as
consisting of rings (r), intermediate types (rs) and spiral-like
features (s). The r and s types of structures occur approximately
with equal frequency in early-type (Sa and SBa) galaxies, whereas
spiral-like features predominate in late-type spirals. Within each
Hubble stage, and within the accuracy of the statistics derived from the
data in {\it A Revised Shapley-Ames Catalog of Bright Galaxies} (excluding
peculiar and edge-on galaxies), there appears to be no systematic
differences between the luminosity distributions of parent galaxies
having r, rs and s type structures. Furthermore, no significant
differences are seen between the $(B-V)_o$ color distributions of spiral
(r), intermediate (rs) , and ringed (r) galaxies of a given Hubble
stage. Finally, within the accuracy of the statistics provided by
Sandage \& Tammann (1981), there is no difference in the relative
frequency of r and s type structures between galaxies with, and 
without, bars.

\section{CONCLUSIONS}

Surprisingly it has been found that, for galaxies of Hubble stages
in the range a - ab - b - bc, normal and barred galaxies have indistinguishable 
$(B-V)_o$ color distributions. Furthermore, within any given
Hubble stage, the intrinsic color distribution of galaxies is found to
be insensitive to (or independent from) the presence of spiral-like or
ring-like features. Taken at face value, the fact that intrinsic color
of a galaxy does not depend on the presence (or absence) of a bar is
surprising because one might have expected (see Kormendy \& Kennicutt
2004 for a review) that the existence of a bar would rearrange disk gas
resulting in transport of disk gas to small radii where it reaches high
density and plausibly feeds into star formation.  According to Combes (2008) 
gas is driven inwards by bar torques.  The gas angular momentum is taken up 
by the bar which is sufficient to weaken or destroy it.  The history of 
star formation determines the evolution a bar (Combes 2008) and hence 
the present intrinsic color of a galaxy.  One might therefore have expected the presence 
(or absence) of a bar to have affected present galaxy colors.  Perhaps the apparent 
independence of the intrinsic colors of spirals from the presence (or absence) of 
bars hints at the possibility that some bars could be ephemeral structures. 
Since it is not the central mass concentration which destroys the bar, it is relatively 
easy to reform a bar after a bar episode is completed.  With significant cosomological 
gas accretion rates several bar episodes might occur in a galaxy with timescales 
of a few Gyr.  On the other hand S$\rm \acute{a}$nchez-Bl$\rm \acute{a}$zquez et al. (2011) 
have argued that some bars formed long ago and have survived to the present day.

I am indebted to Ron Buta, John Kormendy and Preethi Nair for a
number of exchanges of e-mail regarding the differences between normal
and barred spiral galaxies. I also thank Jason Shrivell and Brenda
Parrish for technical assistance and the referee for a number of helpful 
suggestions.

\clearpage

\begin{deluxetable}{lcccc}
\tablewidth{0pt} 

\tablecaption{Median Colors of Normal and Barred Spiral Galaxies}

\tablehead{\colhead{Type}  & \colhead{$(B-V)_{o}^{*}$}   & \colhead{n} & \colhead{25\%}  & \colhead{75\%}} \hspace*{\fill} \\

\startdata

Sa      &  0.82 +/- 0.01    &   41     & 0.79  & 0.86\\
SBa     &  0.84 ~~~ 0.02    &   24     & 0.81  & 0.92\\
\hspace*{\fill} \\

Sab     &  0.74 ~~~0.02     &  32    &  0.67  &  0.81\\
SBab    &  0.77 ~~~0.09     &  5     &  0.67: &  0.87:\\
\hspace*{\fill} \\

Sb      &  0.68 ~~~0.01     &  75    &  0.64  & 0.76\\
SBb     &  0.68 ~~~0.03     &  30    &  0.54  & 0.72\\
\hspace*{\fill} \\

Sbc     &  0.60 ~~~0.02     &  54    &  0.55  & 0.67\\
SBbc    &  0.62 ~~~0.02     &  38    &  0.57  & 0.67\\
\hspace*{\fill} \\

Sc      &  0.53 ~~~0.01     &  188   &  0.44   & 0.57\\
SBc     &  0.48 ~~~0.02     &   45   &  0.44   & 0.58\\
\hline
\multicolumn{4}{l}{Uncertain values are marked by a colon}

\enddata

\end{deluxetable}

\clearpage

\begin{deluxetable}{lcccc}
\tablewidth{0pt}   
\tablecaption{Median Luminosities of Normal and Barred Galaxies}
\tablehead{\colhead{Type}  & \colhead{$M_{B}^{*}$}   & \colhead{n} & \colhead{25\%}  &  \colhead{75\%}}\hspace*{\fill} \\

\startdata
E   & -21.32 +/- 0.11    & 144   & -21.85  &  -20.58\\
\hspace*{\fill} \\
S0  & -20.44 ~~~0.12     &  99   & -21.11  & -19.96\\
SB0 & -20.33 ~~~0.18     &  31   & -20.76  & -19.77\\
\hspace*{\fill} \\
Sa  & -21.04 ~~~0.13    &  68    & -21.53  & -20.49\\
SBa & -21.04 ~~~0.17    &  29    & -21.24  & -20.35\\
\hspace*{\fill} \\
Sab & -21.10 ~~~0.20    & 36     & -21.95  & -20.73\\
SBab& -22.22 ~~~0.27    &  9     & -22.57  & -21.75\\
\hspace*{\fill} \\
Sb  & -21.65 ~~~0.17    & 91     & -22.53   & -20.88\\
SBb & -21.81 ~~~0.18    & 37     & -22.35   & -21.27\\
\hspace*{\fill} \\
Sbc  & -21.30 ~~~0.13  & 68      & -21.83   & -20.75\\
SBbc & -21.41 ~~~0.16  & 47      & -22.04   & -20.91\\
\hspace*{\fill} \\
Sc  &  -20.86~~~0.09  & 264      & -21.56   & -20.99\\
SBc & - 20.20~~~0.17  & 69       & -21.06   & -19.61\\
\enddata

\end{deluxetable}

\begin{deluxetable}{lcccc}
\tablewidth{0pt}   
\tablecaption{Median luminosities of Normal and Barred Galaxies for
objects located in clusters or with $V > 2000~km s^{-1}$}

\tablehead{\colhead{Type}  & \colhead{$M_{B}^{*}$}   & \colhead{n} & \colhead{25\%}  &  \colhead{75\%}}\hspace*{\fill}\\

\startdata

E   &   -21.61  +/- 0.09  & 99   & -22.08  & -21.16 \\
\hspace*{\fill} \\
S0  &   -20.95 ~~~~0.23   & 50   & -21.48  & -19.82\\
SB0 &   -20.30 ~~~~0.38   & 17   & -21.33  & -19.77\\
\hspace*{\fill} \\

Sa  &   -21.44~~~~~0.17   & 37   & -22.08 & -21.04\\
SBa &   -21.18~~~~~0.12   &  12  & -21.33 & -20.90\\
\hspace*{\fill} \\

Sab &   -21.85~~~~0.30    & 18   & -22.42 & -21.17\\
SBab &  -21.96~~~~0.25    &  6    & -22.57 & -21.96\\
\hspace*{\fill} \\

Sb  &   -22.33~~~~0.16    &  41    & -22.69 & -21.65\\
SBb &   -21.92~~~~0.15    &  25    & -22.48 & -21.74\\
\hspace*{\fill} \\

Sbc &   -21.79~~~~0.17    &  34    & -22.33  & -21.31\\
SBbc &  -21.71~~~~0.17    &  26    & -22.23  & -21.37\\
\hspace*{\fill} \\

Sc  &   -21.52~~~~0.09    &  102   & -21.93  & -20.98\\
SBc &   -21.05~~~~0.36    &  24    & -21.66  & -19.89\\

\enddata
\end{deluxetable}

\clearpage

\begin{deluxetable}{lccc}
\tablewidth{0pt}   
\tablecaption{Luminosity Distribution of Early-type Galaxies}

\tablehead{\colhead{$M_B$}  & \colhead{E}   & \colhead{S0+SB0} &  \colhead{Sa+SBa}}

\startdata

-23.00 to -23.49 &  1   &    0   &     0\\
-22.50 to -22.99 & 12   &    1   &     2\\
-22.00 to -22.49 & 13   &    3   &     8\\
-21.50 to -21.99 & 36   &   11   &    13\\
-21.00 to -21.49 & 29   &   19   &    32\\
-20.50 to -20.99 & 20   &   21   &    14\\
-20.00 to -20.49 & 11   &   21   &    15\\
-19.50 to -19.99 & 12   &   30   &    10\\
-19.00 to -19.49 &  4   &   16   &     2\\
-18.50 to -18.99 &  3   &    4   &     1\\
-18.00 to -18.49 &  1   &    3   &     0\\
-17.50 to -17.99 &  0   &    0   &     0\\
-17.00 to -17.49 &  1   &    1   &     0\\
     $ >$ -17.00   &  1   &    0   &     0\\

Total            & 144  &   130  &     97\\

\enddata
\end{deluxetable}

\begin{figure}
\plotone{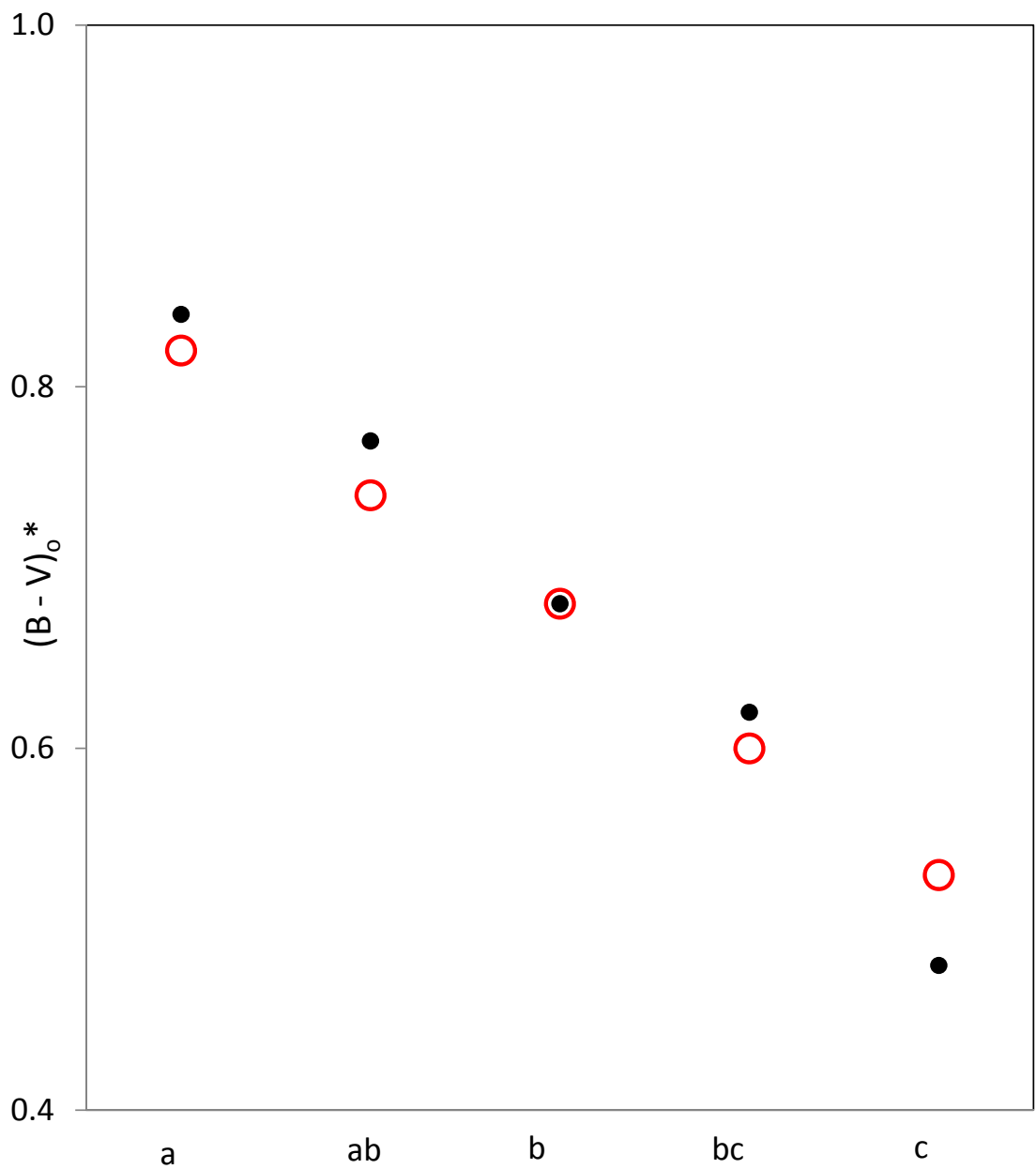}

\caption{Intrinsic colors of normal spirals (red circles ) and 
of barred spirals (blue dots) as a function of Hubble stage. 
The figure shows that the colors of barred and unbarred spirals 
are indistinguishable over the range of Hubble stages 
a - ab - b - bc.}
\end{figure}

\begin{figure}

\plotone{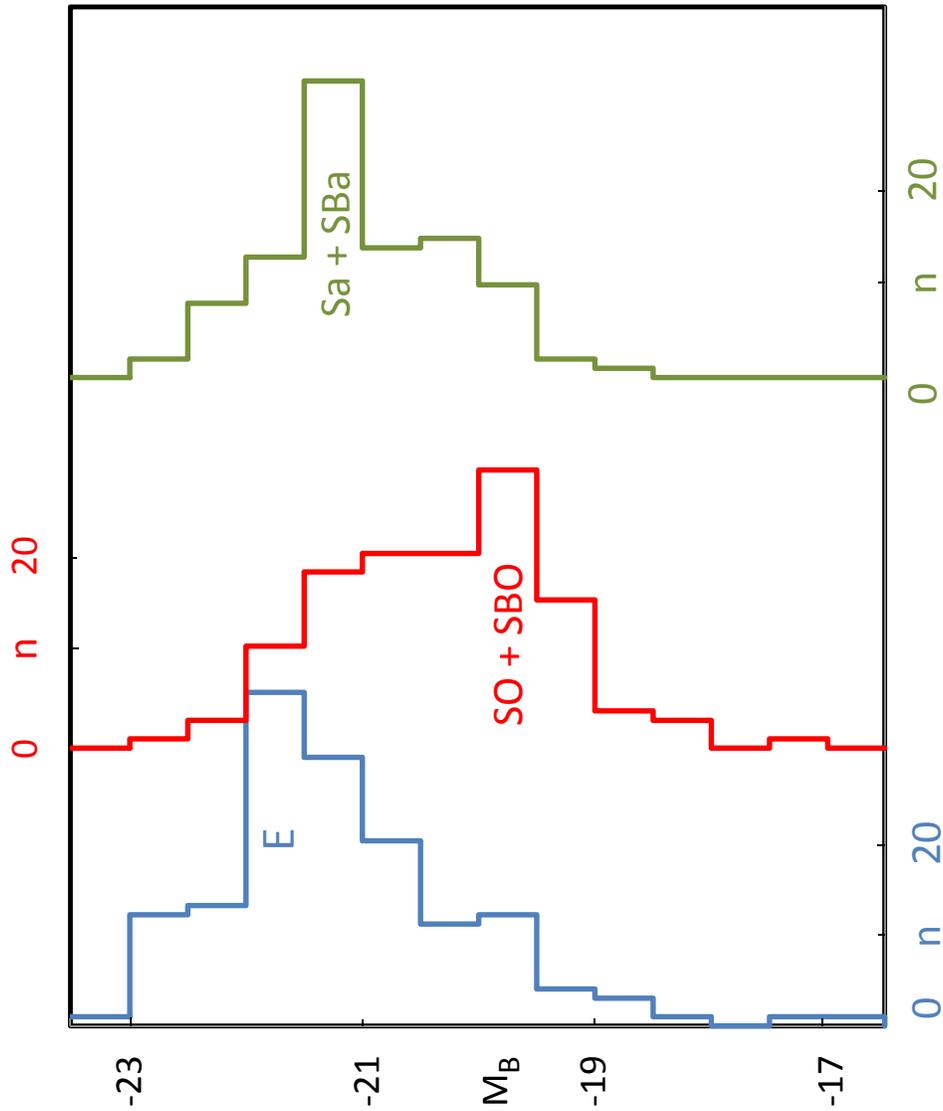}

\caption{ Histogram of the luminosity distribution of elliptical
(red), S0+SB0 (blue) and Sa+SBa galaxies (green) in the
Shapley-Ames catalog. The figure shows that S0 galaxies are
systematically fainter than either E or Sa galaxies. This
suggests that lenticular galaxies are not truly intermediate
between ellipticals and spirals.}
\end{figure}

\end{document}